\documentclass[final,1p,times,authoryear]{elsarticle}
\usepackage{amssymb}
\usepackage{amsmath}
\usepackage{fancyvrb} 
\usepackage{tcolorbox} % to create colored boxes
\usepackage{array}
\usepackage[authoryear]{natbib}

\journal{Journal of Systems and Software}

\begin{document}

\begin{frontmatter}



\title{AlgoTouch: An Execution-Centered Approach to Incremental Construction of Imperative Programs} %% Article title


%% Title
%% \title{AlgoTouch: An Execution-Centered Approach to Incremental Construction of Imperative Programs}

%% Authors
\author[aff1]{Michel Adam}
\ead{michel.jacques.adam@gmail.com}

\author[aff2,aff3]{Patrice Frison}
\ead{patrice.frison@irisa.fr}

\author[aff2]{Moncef Daoud}
\ead{moncef.daoud@univ-ubs.fr}

\author[aff2]{Sabine Letellier Zarshenas}
\ead{sabine.letellier@univ-ubs.fr}

%% Affiliations
\affiliation[aff1]{
  organization={Doctor of Computer Science},
  city={Vannes},
  country={France}
}

\affiliation[aff2]{
  organization={University of South Brittany},
  city={Vannes},
  country={France}
}

\affiliation[aff3]{
  organization={IRISA},
  city={Vannes},
  country={France}
}

%% Abstract
\begin{abstract}
Program construction in imperative languages remains largely based on writing textual code that specifies sequences of instructions operating on program data. This approach requires developers to anticipate the effects of instructions on evolving data states, which increases cognitive load and the likelihood of errors during early and incremental development.

This paper presents AlgoTouch, an execution-based system for incremental construction of imperative programs through direct manipulation of program data. Rather than assembling syntactic structures, programs are constructed by executing concrete data transformations that are recorded and incorporated into an internal intermediate representation. AlgoTouch relies on an explicit notional machine that exposes data storage, computation, and control flow, enabling continuous alignment between observed execution and program structure.

A central contribution of the system lies in its deterministic synthesis of control structures from execution behavior. Conditional statements are derived from observed comparisons, while iterative behaviors are encapsulated in loop macros that support non-linear and incremental construction. This design enables partial and incomplete programs to be executed, refined, and completed while preserving semantic consistency.

AlgoTouch automatically generates correct and readable programs in several mainstream imperative languages, including Python, C, C++, and Java. The system is evaluated through engineering-level validation on a representative set of algorithmic benchmarks, demonstrating correctness, expressiveness, robustness, and language independence.

By integrating execution, construction, and code generation within a unified architecture, this work introduces an alternative model for interactive program construction and contributes a new class of execution-centered development systems.

\end{abstract}


%\item AlgoTouch introduces an execution-centered environment for incremental %program construction.
%\item Program behavior is directly manipulated through data-oriented %interactions.
%\item The approach supports progressive refinement of imperative programs.
%\item The system is validated as a software engineering artifact.

%% Keywords
%% Keywords

\end{frontmatter}



\section{Introduction}
The construction of imperative programs remains a central activity in software engineering. Despite the maturity of programming languages and development environments, constructing correct programs still requires developers to mentally bridge several levels of abstraction: data state, control flow, and executable code. This gap is particularly visible during early stages of development, incremental refinement, and exploratory programming, where programs are often incomplete and subject to frequent modification.

Most programming environments emphasize textual code as the primary representation of programs. While effective for expressing finalized solutions, textual code offers limited support for observing intermediate execution states or for constructing control structures incrementally from concrete behavior. As a result, developers must mentally simulate execution and anticipate control flow, increasing cognitive load and the likelihood of errors.

Several approaches have attempted to address these limitations, including visual programming environments, program synthesis techniques, and execution-based debugging tools. However, many of these approaches either restrict expressiveness, require complete specifications, or treat execution and construction as separate phases. In contrast, modern software engineering increasingly values tools that support incremental construction, partial execution, and continuous validation of program behavior.

This paper introduces AlgoTouch, a software system for interactive program construction based on direct manipulation of data and observed execution. AlgoTouch enables users to build imperative programs by performing concrete data transformations, from which executable code is automatically synthesized. Program construction and execution are tightly coupled: every manipulation both transforms data and contributes to the program structure.

At the core of AlgoTouch lies a notional machine that supports partial execution, incremental definition of control structures, and deterministic code generation. Unlike traditional environments, AlgoTouch allows programs to remain structurally incomplete while preserving internal consistency. Control structures such as conditionals and loops are synthesized progressively from observed execution paths rather than predefined textual constructs.

AlgoTouch generates executable code in multiple imperative languages, including Python, C, C++, and Java, through a common intermediate representation. This design allows the system to remain language-independent while ensuring that generated programs preserve the semantics of observed manipulations.

The contributions of this paper are:
\begin{itemize}
    \item A program construction architecture that integrates direct data manipulation, partial execution, and automatic code generation within a unified system.
    \item A generic mechanism for synthesizing control structures, including conditionals and loops, from observed execution paths.
    \item A language-independent intermediate representation enabling multi-language code generation with preserved semantics.
    \item An engineering-level evaluation demonstrating correctness, expressiveness, robustness, and applicability across a representative set of imperative programs.
\end{itemize}

The remainder of the paper is organized as follows. Section~2 reviews related work on
program construction, execution-based approaches, and direct manipulation systems.
Section~3 presents AlgoTouch and details how imperative programs are generated through
execution and direct manipulation of program data. Section~4 discusses the evolution of
the system and the design trade-offs that shaped its underlying model and interaction
mechanisms. Section~5 evaluates AlgoTouch through representative examples and usage
scenarios. Section~6 concludes the paper and outlines perspectives for future work.

\section{Related works}
Although early work on programming through data manipulation dates back to the 1970s, relatively few systems have explored this approach in a systematic manner. Existing work can be broadly grouped into execution-based program construction systems, programming by demonstration approaches, and direct manipulation environments for code generation. This section reviews representative systems in these areas and positions AlgoTouch with respect to their capabilities and limitations.

From a design and interaction perspective, the importance of making program 
behavior explicit, inspectable, and manipulable during execution has been 
strongly advocated, notably in Bret Victor’s influential essay on learnable 
programming \cite{victor2012learnable}.

\subsection{Programming by Demonstration and Early Execution-Based Systems}

The Pygmalion system \cite{smith1975pygmalion}, is one of the earliest examples of execution-based program construction. Pygmalion allowed users to manipulate graphical objects directly, recording these manipulations as executable programs. The system was Turing-complete and supported conditionals, loops, and recursion. Its treatment of conditionals was particularly innovative: the true branch was generated immediately, while the alternative branch remained undefined until a complementary execution path was observed. AlgoTouch adopts a similar principle for incremental construction of conditional structures.

SmallStar \cite{halbert1984programming} extended Pygmalion’s ideas within the Xerox Star environment by separating object manipulation from code visualization. Programs were generated exclusively from recorded interactions, with explicit mechanisms for inserting, deleting, and executing recorded instructions. Control structures were constructed by encapsulating recorded actions within predicates or iteration constructs. Although powerful, SmallStar relied on explicit recording phases and offered limited support for partial or non-linear program construction.

The macro system of the Emacs editor \cite{stallman1993gnu} provides a lightweight form of programming by demonstration for automating repetitive tasks. Macros record sequences of operations that can later be replayed or edited textually. Unlike Pygmalion and SmallStar, Emacs macros do not provide an explicit execution model for incremental construction of control flow. Nevertheless, AlgoTouch borrows the notion of macros as reusable execution units from this approach.

\subsection{Execution-Aware and Live Programming Systems}

Execution-aware environments integrate program execution into the development process by exposing runtime state alongside program construction. Alvis Live! \cite{hundhausen2009can} provides real-time visualization of program variables during execution and introduces explicit index variables for array traversal. While Alvis Live! supports limited generation of code from data manipulation, it restricts control structures to array traversal loops and 
simple decision structures without an {\tt else} clause and does not support construction of Turing-complete programs through manipulation alone. Its primary focus remains execution visualization rather than program synthesis.

AlgoTouch similarly exposes runtime state during construction but differs by treating execution as the primary mechanism for defining program structure. Control flow and data dependencies are derived incrementally from observed execution paths rather than specified explicitly.

\subsection{Direct Manipulation for Code Generation}

CodeInk \cite{scott2014direct} allows users to manipulate data structures to generate Python code, supporting assignments, insertions, and comparisons. However, comparisons are recorded only as comments, preventing the generation of executable conditionals and loops. As a result, CodeInk cannot produce Turing-complete programs through manipulation alone and offers limited support for control-flow synthesis.

ManipoSynth \cite{hempel2022maniposynth} adopts a bimodal approach in which users may alternate between textual programming and direct manipulation of values. Programs are constructed non-linearly and may contain undefined parts that are completed later. While ManipoSynth supports synthesis from examples and partial programs, it is primarily oriented toward functional programming and does not emphasize execution-based construction of imperative control flow.

Sketch-n-Sketch \cite{chugh2016programmatic}, \cite{hempel2019sketch} 
and Twoville \cite{johnson2023computational} explore bidirectional programming for graphics, allowing users to manipulate visual representations and synchronize changes with underlying code. These systems focus on graphical objects rather than program variables and do not address construction of general-purpose imperative programs from execution semantics.

Within execution-based approaches to program construction, some systems adopt a data-centered
perspective in which programs are built by reasoning about variable states during execution.

\subsection{Data-Centered Program Construction Systems}

AlgoT \cite{thorgeirsson2024comparing} constructs programs incrementally by manipulating program variables and selecting operations based on current data state. Code is generated instruction by instruction in a restricted algorithmic language that does not support iterative constructs directly. Execution traces can be navigated post hoc to inspect variable states. Although AlgoT is data-centered, its construction model differs fundamentally from AlgoTouch: control flow is specified explicitly and repetition is handled through recursion rather than synthesized from execution.

\subsection{Summary and Positioning}

Existing systems demonstrate the feasibility of constructing programs from execution and data manipulation, but they exhibit important limitations: restricted control structures, lack of support for partial programs or tight coupling to specific languages or domains.

AlgoTouch differs from prior work by combining:
\begin{itemize}
    \item deterministic interpretation of execution steps,
    \item incremental synthesis of conditionals and loops from observed execution paths,
    \item explicit support for incomplete programs and partial execution,
    \item language-independent intermediate representation enabling multi-language code generation.
\end{itemize}

This combination positions AlgoTouch as an execution-based program construction system rather than a visual programming environment or a programming-by-demonstration tool.

\section[Code generation by direct manipulation]{Code Generation by Direct Manipulation}
\label{sec:code_generation}

This section describes how AlgoTouch supports the construction of Turing-complete imperative programs exclusively through execution and direct manipulation of program data. The proposed model integrates a simple abstract machine, a language-independent intermediate representation (called AGT) and an execution-centered interaction model that enables deterministic code generation from concrete data transformations.

The design follows a classical separation between algorithm, machine model, and programming language. Program construction proceeds by executing data transformations on an explicit abstract machine. These transformations are recorded, structured into control constructs, and translated into executable programs in multiple target languages.

We first describe the underlying machine model, the system interface, and the supported data abstractions. We then detail the mechanisms for operation recording, macro construction, synthesis of conditionals and loops, macro composition, and execution modes.

\subsection[Machine Model]{Machine Model}
\label{sec:machine}

AlgoTouch relies on an explicit stored-program abstract machine that defines the operational semantics of all generated programs. The machine consists of:

\begin{itemize}
\item a memory composed of addressable locations storing integers or characters,
\item a processor supporting arithmetic, comparison, and assignment operations,
\item a single sequential control thread with conditional branching and non-recursive procedure calls,
\item input and output devices connected directly to memory.
\end{itemize}

The instruction set comprises four categories:

\begin{itemize}
\item memory access (read and write),
\item arithmetic and comparison operations,
\item input/output transfers,
\item control-flow operations (conditional branch, call, return).
\end{itemize}

This machine model provides a deterministic execution semantics that underlies both construction and execution. All manipulations performed in the interface correspond to concrete machine-level actions and are mapped unambiguously to the internal intermediate representation.

\subsection[Interface Overview]{Interface Overview}
\label{sec:interface}

\begin{figure}[htb]
\centering
\includegraphics[width=\textwidth]{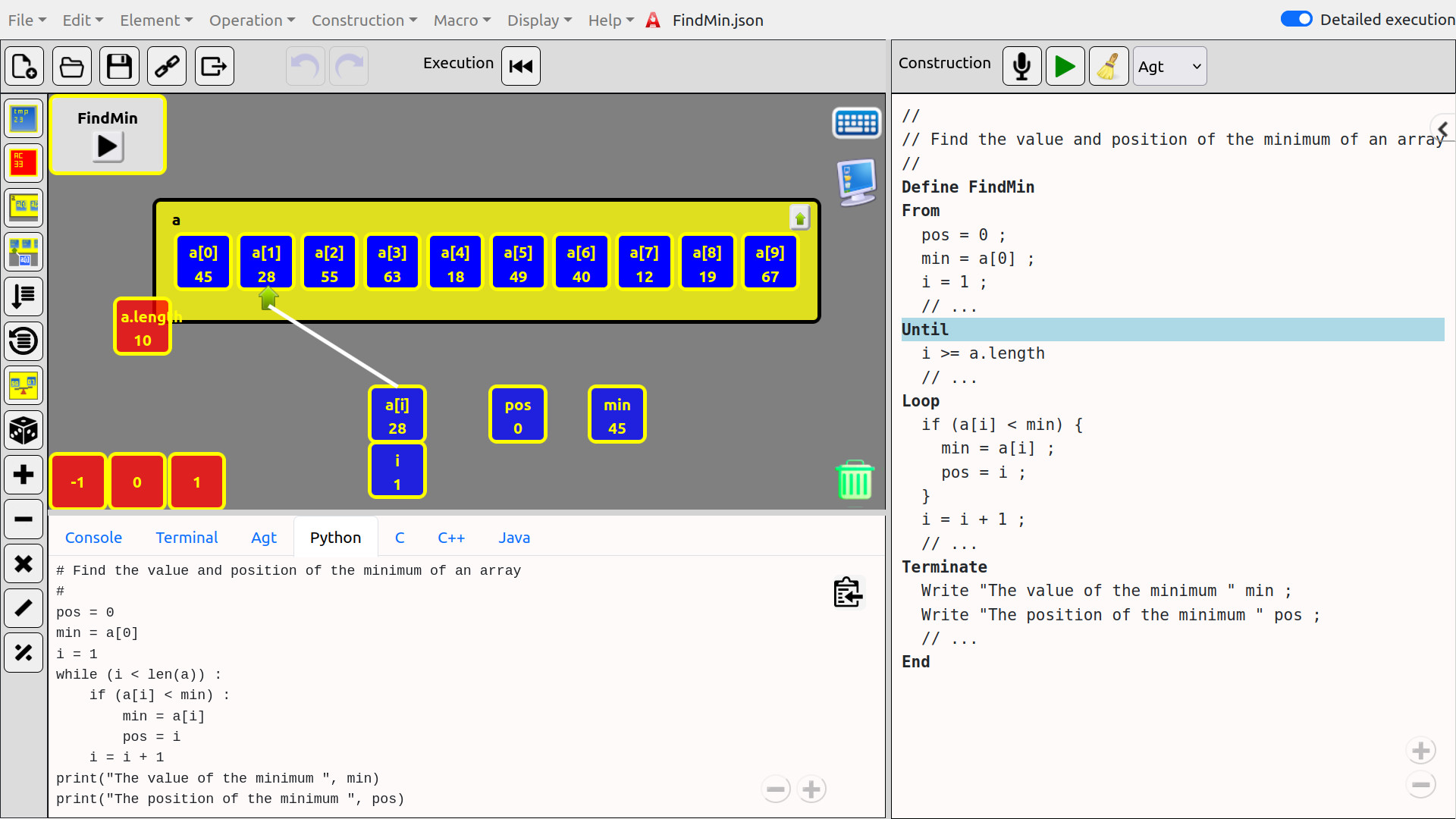}
\caption{ 
The AlgoTouch interface showing a program under construction, called {\tt FindMin}, for searching an array.
It displays the smallest value and its position.
The Workspace, in the center, presents an array {\tt a} of 10 elements indexed by the variable {\tt i}, and two variables {\tt pos} and {\tt min}.
The AGT program is shown on the right, and the Python version is available at the bottom.}
\label{fig:interface}
\end{figure}

AlgoTouch exposes program construction through a unified workspace that displays program data, available operations, and generated code simultaneously (Figure~\ref{fig:interface}).
The interface is organized into:

\begin{itemize}
\item a \textbf{Workspace} showing variables, arrays, indexes, and macros as explicit data objects,
\item a \textbf{Sidebar} for creating data objects and operations,
\item an \textbf{Instructions Area} displaying the generated program in the AGT intermediate language and in target languages,
\item a \textbf{Construction Area} controlling recording, execution, deletion, and language selection,
\item a \textbf{Console and Export Area} presenting execution traces and translated programs,
\item standard toolbars and menus.
\end{itemize}

Although programs are never edited textually, the generated code is continuously visible and synchronized with execution, providing an explicit correspondence between behavior and structure.

\subsection[Data Model]{Data Model}
\label{sec:data}

AlgoTouch supports a minimal yet expressive set of data abstractions:

\begin{itemize}
\item scalar variables and constants,
\item one-dimensional arrays,
\item index variables bound to arrays.
\end{itemize}

All data objects are persistent and globally visible throughout construction and execution. Index variables provide an explicit representation of array access and are visually bound to the referenced element. Out-of-bounds accesses are detected dynamically and signaled immediately.

This explicit representation, simular to Alvis Live! in \cite{hundhausen2009can}, enables deterministic tracking of data dependencies and supports automatic generation of safe array access code.

\subsection[Operations and Recording]{Operations and Recording}
\label{sec:operations}

Program construction is performed by executing elementary operations on data objects in the workspace, through direct manipulation. Operations are specified by selecting, dragging, and dropping variables, constants, and array elements onto operation operators, thereby expressing assignments, arithmetic transformations, comparisons, and input/output transfers directly on the program state.

Each manipulation is interpreted as a concrete machine operation and immediately executed on the underlying abstract machine. When recording mode is active, the executed operation is simultaneously captured and translated into the internal intermediate representation. This establishes a strict correspondence between the performed manipulation, the resulting data state, and the generated instruction.

Because operations are expressed as explicit data transformations rather than textual commands, construction proceeds through observable state changes. The resulting representation remains independent of target language syntax and serves as the basis for deterministic multi-language code generation.

\subsection[Macro-Based Program Construction]{Macro-Based Program Construction}
\label{sec:macros}

Program structure in AlgoTouch is organized around \emph{macros}, which represent reusable execution blocks synthesized exclusively from sequences of data manipulations.

Two macro types are supported:

\begin{itemize}
\item \textbf{Simple Macros}: straight-line instruction sequences, including conditionals, obtained by recording successive data transformations,
\item \textbf{Loop Macros}: structured iterative constructs synthesized from repeated executions on evolving data states.
\end{itemize}

A macro is created empty and populated by recording executed operations performed through direct manipulation of data objects. At no point are instructions entered textually: the macro body is entirely inferred from observed data transformations and corresponding machine operations.

Macros can be executed, composed, and invoked as atomic instructions. This abstraction provides modular construction, hierarchical control composition, and support for both bottom-up and top-down development while preserving a strict execution-based semantics.

\subsection[Synthesis of Conditional Structures]{Synthesis of Conditional Structures}
\label{sec:conditional}

Sequential execution yields linear code directly. Conditional structures require synthesis of divergent control flow from concrete execution behavior observed during data manipulation.

When a comparison is introduced through direct manipulation of data values, AlgoTouch generates a conditional construct whose guard is selected among predicates that hold in the current data state. Construction proceeds by defining first the branch corresponding to the current execution path, leaving the alternative branch temporarily undefined. In the generated program, this undefined branch is explicitly marked by a \texttt{TODO} placeholder indicating an incomplete control region.

To specify the missing branch, execution is restarted with a data configuration that inverts the predicate outcome. When control reaches the \texttt{TODO}-marked branch, execution halts and recording resumes, allowing the user to complete the branch by performing the required data manipulations. The corresponding instructions are then inferred from these transformations and replace the placeholder region.

This mechanism ensures that each branch is specified under an executable data state, memory consistency is preserved, and both branches remain semantically aligned with observed behavior. The resulting conditional is deterministic and language-independent.

\paragraph{Illustrative example.}

We illustrate the construction of a conditional structure on a simple but
representative example.  
The program modifies a variable {\tt v} according to its sign: if {\tt v} is
negative or zero, it is incremented; otherwise, it is decremented.

The user first establishes a concrete data state by directly manipulating the
variable {\tt v} in the Workspace (e.g., by drag-and-drop assignment) so that
{\tt v~=~-3}.  
Recording mode is then activated and a comparator is introduced by selecting
{\tt v} and dragging it onto the comparison operator, together with the
constant {\tt 0}.  
Among the relations that evaluate to true in the current state, the user selects
{\tt v~<=~0}, which inserts the conditional structure into the program.

Still in the same data configuration, the user performs the manipulation
corresponding to this case by incrementing {\tt v} through a left-right swiping gesture.  
This generates the instruction {\tt v~=~v~+~1}, which is recorded as the body of
the true branch.  
At this point, the alternative branch has not yet been specified and is
explicitly marked in the generated AGT code by a placeholder:
\begin{tcolorbox}[colframe=black, colback=white, boxrule=0.5pt]
\begin{Verbatim}[samepage=true]
    if (v <= 0) {
        v = v + 1 ;
    } else {
        // TODO : v > 0
    }
\end{Verbatim}
\end{tcolorbox}

To define the missing branch, the user establishes a different data state by
assigning a positive value to {\tt v} (e.g., {\tt v~=~1}) with Recording mode
disabled, and restarts execution in Construction mode.  
When execution reaches the conditional, control flows into the undefined branch,
and execution is suspended, indicating that code remains to be completed.

Recording mode is automatically reactivated.  
The user then performs the appropriate data manipulation by decrementing
{\tt v}, which generates the instruction {\tt v~=~v~-~1} and replaces the
{\tt TODO} placeholder.  
The completed conditional becomes:
\begin{tcolorbox}[colframe=black, colback=white, boxrule=0.5pt]
\begin{Verbatim}[samepage=true]
    if (v <= 0) {
        v = v + 1 ;
    } else {
        v = v - 1 ;
    }
\end{Verbatim}
\end{tcolorbox}

This example illustrates how conditional code is constructed by successively
instantiating concrete data states and manipulating program data directly.
Each branch is specified in a configuration that makes it executable, preserving
a tight correspondence between memory state, user action, and generated code.

\subsection[Loop Macros]{Loop Macros}
\label{sec:macro_loop}

Loops are synthesized using a single generic construct inspired by the Eiffel language in  \cite{meyer1992eiffel} . A Loop Macro consists of four blocks:

\begin{itemize}
\item \textbf{From}: initialization,
\item \textbf{Until}: list of exit conditions,
\item \textbf{Loop}: repeated body,
\item \textbf{Terminate}: post-loop actions.
\end{itemize}

Exit conditions are specified as a list of simple comparisons rather than a single Boolean expression. Conditions are evaluated sequentially and execution exits as soon as one holds.

This design simplifies termination reasoning, eliminates compound Boolean expressions, and enables automatic ordering of safety conditions such as index bounds before array access.

\subsection[Methodology for Loop Construction]{Methodology for Loop Construction}
\label{sec:method}

Loop construction follows a non-linear methodology exploiting partial execution:

\begin{enumerate}
\item Define the loop body by executing a representative iteration.
\item Execute repeated iterations until a stable or terminating configuration is reached.
\item Generate exit conditions from the observed termination state.
\item Add initialization and boundary conditions incrementally.
\end{enumerate}

This process allows exit conditions to emerge naturally from concrete execution behavior rather than being specified a priori.

\paragraph[Example: Insertion with Multiple Exit Conditions]
{Example: Insertion with Multiple Exit Conditions}
\label{sec:loop_insertion}

We illustrate the loop construction methodology using the inner loop of insertion sort, a
representative algorithme that naturally requires multiple exit conditions.

Assume that the sub-array \texttt{a[0..i-1]} is already sorted and that the value stored at
position \texttt{i} must be inserted into its correct position. The algorithm repeatedly
exchanges adjacent elements, moving the current value leftward until either its correct
position is reached or the beginning of the array is encountered.%
\footnote{Insertion sort is used here purely for illustrative purposes. Its step-by-step
behavior, local data transformations, and simple control structure make it particularly
well suited to demonstrate execution-based program construction, without loss of
generality.}

All components of the loop are specified exclusively through direct data manipulation; no
textual instruction is written by the user.

\paragraph{Loop body construction}
The construction process starts with the definition of a representative iteration.
The user establishes a concrete data configuration corresponding to a generic execution
step. Two index variables are introduced: \texttt{k}, initially equal to \texttt{i},
designating the element being inserted, and \texttt{j}, equal to \texttt{k-1}, designating
its left neighbor.

Using drag-and-drop manipulation on indexed array elements, the user performs an
elementary swap between \texttt{a[k]} and \texttt{a[j]}, followed by decrements of
\texttt{j} and \texttt{k}. These manipulations are recorded in the \texttt{Loop} block and
generate the following code:

\begin{tcolorbox}[colframe=black, colback=white, boxrule=0.5pt]
\begin{Verbatim}[samepage=true]
Loop
    tmp = a[k] ;
    a[k] = a[j] ;
    a[j] = tmp ;
    j = j - 1 ;
    k = k - 1 ;
\end{Verbatim}
\end{tcolorbox}
At this stage, the loop body captures a generic iteration independently of any termination
condition.

\paragraph{Derivation of exit conditions}
In AlgoTouch, exit conditions are not inferred automatically from abstract loop schemas.
Instead, the user explicitly constructs, through data manipulation, representative data
states corresponding to each scenario in which the loop should terminate.

Two distinct exit scenarios arise in this example:

\begin{itemize}
\item The element reaches a position where it is greater than or equal to its predecessor.
\item The element reaches the beginning of the array (index~0).
\end{itemize}

To capture the first scenario, execution in Construction mode is continued through repeated
manipulation-driven iterations until a stable configuration is reached in which no further
swap is required. From this concrete data state, the user records the comparison
\texttt{a[j]~<=~a[k]}, which becomes the first exit condition.

To capture the second scenario, the user positions the data so that the element is moved all
the way to index~0. When further swapping would cause an invalid access, the index
\texttt{j} becomes negative and is visually highlighted. From this configuration, the user
records the exit condition \texttt{j~<~0}.

The resulting exit block is:
\begin{tcolorbox}[colframe=black, colback=white, boxrule=0.5pt]
\begin{Verbatim}[samepage=true]
Until
    j < 0
    a[j] <= a[k]
\end{Verbatim}
\end{tcolorbox}

Both exit conditions are thus derived directly from concrete execution states explicitly
constructed by the user. AlgoTouch automatically orders the conditions so that index bounds
are tested before array access, preventing out-of-bounds errors without requiring manual
reasoning about evaluation order.

\paragraph{Loop initialization.}
The loop initialization cannot be fully defined until the complete loop body has been
constructed. Indeed, the initialization must establish a data configuration that is
consistent with the operations performed during a generic iteration, in particular with
the index variables and data dependencies introduced in the \texttt{Loop} block.

Once the structure of the loop body is known, the user specifies the initialization by
selecting the \texttt{From} block and enabling Recording mode. In this example, the user
copies \texttt{i} into \texttt{k} and \texttt{i~-~1} into \texttt{j}.

This operation assigns \texttt{j} and \texttt{k} the same initial values that were
previously established manually with Recording mode disabled, but now records them
explicitly as part of the loop definition. The resulting initialization code is:

\begin{tcolorbox}[colframe=black, colback=white, boxrule=0.5pt]
\begin{Verbatim}[samepage=true]
From
    k = i ;
    j = i - 1 ;
\end{Verbatim}
\end{tcolorbox}

Although the initialization is presented here after the derivation of all exit
conditions, it does not conceptually depend on their complete specification. In practice,
the \texttt{From} block may be generated as soon as the loop body is known, even if some
exit conditions remain to be identified. This flexibility reflects the non-linear nature
of loop construction in AlgoTouch.

\paragraph{Resulting loop macro}
After defining the loop body, exit conditions, and initialization, the loop macro is fully
specified. In this example, the \texttt{Terminate} part of the macro is empty, as no
additional actions are required upon loop termination.

The final code synthesized by AlgoTouch for this example is shown below:

\begin{tcolorbox}[colframe=black, colback=white, boxrule=0.5pt]
\begin{Verbatim}[samepage=true]
Define InsertElt

From
    k = i ;
    j = i - 1 ;

Until
    j < 0
    a[j] <= a[k]

Loop
    tmp = a[k] ;
    a[k] = a[j] ;
    a[j] = tmp ;
    j = j - 1 ;
    k = k - 1 ;

Terminate

End
\end{Verbatim}
\end{tcolorbox}
This loop macro inserts the element \texttt{a[i]} into the already sorted prefix
\texttt{a[0..i-1]} by iteratively swapping adjacent elements until one of the exit
conditions is satisfied. The macro structure makes explicit the separation between
initialization, iteration, termination conditions, and post-loop behavior, while remaining
entirely derived from observed execution states.

\subsection[Macro Composition and Program Construction]{Macro Composition and Program Construction}
\label{sec:composition}

Complex programs in AlgoTouch are constructed by composing macros through macro calls.
Adding a macro call to the program is achieved by executing the corresponding macro in
recording mode. This composition mechanism supports both bottom-up and top-down development
strategies, allowing programmers to structure programs incrementally while maintaining a
close connection between execution and specification.

For example, the \texttt{InsertionSort} macro shown below illustrates how a loop macro
invokes another macro (\texttt{InsertElt}) within its body. The macro call is introduced by
executing \texttt{InsertElt} in recording mode, resulting in a structured loop whose control
flow is synthesized from execution.
\begin{tcolorbox}[colframe=black, colback=white, boxrule=0.5pt]
\begin{Verbatim}[samepage=true]
Define InsertionSort
    From
        i = 1 ;
    Until
        i >= a.length
    Loop
        InsertElt;
        i = i + 1 ;
        Terminate
End
\end{Verbatim}
\end{tcolorbox}

\paragraph{Bottom-up composition}
In bottom-up composition, reusable macros are defined first and subsequently invoked as
atomic instructions within higher-level macros. Each macro encapsulates a validated
behavior and can be reused without modification.

This approach is consistent with classical principles of modular program decomposition,
in which independently developed components are combined through well-defined interfaces
(e.g., \cite{Parnas1972}). Nested control structures are expressed through macro calls
rather than through syntactic nesting of loops or conditionals. As a result, each loop or
control structure can be developed, tested, and reasoned about independently before being
integrated into a larger program. This promotes modularity and reuse while preserving the
execution-based nature of program construction.

\paragraph{Top-down construction}
Macro calls may be introduced before the corresponding macro bodies are defined. In this
top-down approach, the user specifies and records the calling macro first, even when some
of its constituent operations are not yet available as executable macros.

When execution reaches a call to an undefined macro, execution is suspended at that point.
The user then produces the intended effect of the missing macro through direct data
manipulation. These manipulation-driven executions act as concrete simulations of the
future macro behavior, allowing the user to explore, refine, and validate the expected
algorithmic effect in context.

Once the calling macro has been fully specified and validated, the user can reuse the data
manipulations performed during these simulations to explicitly construct the body of the
called macro. The macro code is not synthesized automatically from demonstrations; rather,
it is defined by the user, guided by the stabilized execution patterns observed during
interactive manipulation.

This mechanism differs fundamentally from the use of stubs in traditional top-down
program development, as described in classical software engineering literature
(e.g., \cite{Brooks1987,Sommerville2016}). A stub typically provides a provisional
implementation whose role is to satisfy an abstract interface or control-flow requirement,
often returning fixed or simplified values to enable early integration and testing. In
contrast, an undefined macro in AlgoTouch has no predefined behavior: its effect is
produced interactively through concrete data manipulation, within the actual execution
context of the calling macro.

This top-down construction mechanism differs from program synthesis approaches, where
executable code is inferred automatically from examples or specifications
(e.g., \cite{gulwani2017programming, solar-lezama2008sketching}). In AlgoTouch, macro definition
follows a user-driven construction process: data manipulation supports algorithmic
reasoning and design, while the structure and content of the macro remain under explicit
human control.

\subsection[Execution and Visualization]{Execution and Visualization}
\label{sec:execution}

Execution plays a central role in both program construction and validation. AlgoTouch
provides several execution modes:

\begin{itemize}
\item \textbf{Construction mode}: selective execution of individual instructions or blocks
to support incremental completion,
\item \textbf{Direct mode}: full execution of a macro,
\item \textbf{Animation mode}: stepwise visualization of data evolution,
\item \textbf{Detailed / non-detailed modes}: continuous or summarized presentation of
state changes.
\end{itemize}

Construction mode enables the user to execute programs from arbitrary points, complete
missing branches, identify loop termination conditions, and validate partial or evolving
program structures. Execution is driven by the current data configuration rather than by a
predefined control-flow path.

Although this mode supports activities traditionally associated with debugging, it differs
fundamentally from the use of a classical debugger as described in the software engineering
and programming tools literature (e.g., \cite{Myers1979,Zeller2009}). In a debugger,
execution typically starts from the beginning of the program and is interrupted at
predefined breakpoints, with the primary goal of observing, isolating, and diagnosing
erroneous behavior.

In AlgoTouch, no breakpoints are required: the user directly selects the instruction or
block to execute and may redefine the data state before execution. This makes it possible
to test, explore, and validate specific program fragments in isolation, without replaying
the entire execution history, a limitation commonly noted in traditional debugging
workflows (e.g., \cite{Ko2006}).

As a result, execution in AlgoTouch is not merely a post hoc diagnostic activity, but an
integral component of program construction. Construction, execution, visualization, and
validation are tightly interwoven within a unified execution-centered framework.

\subsection[Supporting Mainstream Languages: A Design Challenge]{Supporting Mainstream Languages: A Design Challenge}
\label{sec:mainstream}

A central design objective of AlgoTouch is to support learning and reasoning about
algorithms while remaining compatible with mainstream programming languages. This poses a
non-trivial challenge: AGT introduces high-level, execution-oriented control constructs
(\texttt{From}, \texttt{Until}, \texttt{Loop}, \texttt{Terminate}) that have no direct
syntactic counterparts in conventional languages such as Python, Java, or C.
In AlgoTouch, these mainstream languages can be used directly as supporting languages in
place of AGT. To enable this, AGT control constructs are represented explicitly as
annotated regions in the target language. In particular, the control structure of a macro
is reflected through structured comments (e.g., \texttt{\#~Initialization},
\texttt{\#~Exit conditions}, \texttt{\#~Loop body}, \texttt{\#~Termination}), which play a
role analogous to the corresponding AGT blocks. These comments are selectable and active
within the interface, just like the \texttt{From}, \texttt{Until}, \texttt{Loop}, and
\texttt{Terminate} constructs in AGT.

This correspondence-based approach allows users to work interchangeably in AGT or in a
mainstream language while preserving the execution-centered interaction model.

\paragraph{Example.}
Figure~\ref{tab:AGT_Python} presents a representative example illustrating this approach.
The left column shows the AGT macro \texttt{InsertElt}, while the right column displays the
corresponding Python code generated by AlgoTouch.
\begin{figure}[ht]
\centering
\begin{tcolorbox}[
colback=white, % Background color
colframe=black, % Border color
arc=4pt, % Radius of rounded corners
boxrule=.5pt, % Border thickness
boxsep=5pt, % Inner spacing
left=2pt, right=2pt, top=2pt, bottom=2pt, % Internal margins
width=\textwidth % Width of the box
]
\begin{minipage}[b]{0.49\textwidth}
\begin{tabular}{m{0.9\textwidth}}
\textbf{AGT Program}\\\hline
\begin{Verbatim}
Define InsertElt
    From
        k = i ;
        j = i - 1 ;
    Until
        j < 0
        a[j] <= a[k]
        
        
    Loop
        tmp = a[k] ;
        a[k] = a[j] ;
        a[j] = tmp ;
        j = j - 1 ;
        k = k - 1 ;
    Terminate
End
\end{Verbatim}
\end{tabular}
\end{minipage}
\hfill
\begin{minipage}[b]{0.49\textwidth}
\begin{tabular}{m{0.9\textwidth}}
\textbf{Equivalent in Python}\\\hline
\begin{Verbatim}
# Macro InsertElt
# Initialization
k = i
j = i - 1
# Exit conditions:
# Exit if j < 0
# Exit if a[j] <= a[k]
#
while (j >= 0 and a[j] > a[k]):
    # Loop body
    tmp = a[k]
    a[k] = a[j]
    a[j] = tmp
    j = j - 1
    k = k - 1
# Termination
#
\end{Verbatim}
\end{tabular}
\end{minipage}
\end{tcolorbox}
\caption{Parallel representation of an AGT macro and its Python counterpart.
AGT control blocks are mapped to selectable comments in Python, making their semantic roles
explicit. Exit conditions specified in the \texttt{Until} block are shown individually and
are related to the continuation condition of the \texttt{while} loop, which is formed as the
conjunction of their negations.}
\label{tab:AGT_Python}
\end{figure}

\paragraph{Semantic correspondence rather than translation}
AGT control constructs do not correspond to primitive control structures in Python.
Consequently, the generated code should not be interpreted as a direct syntactic translation.
Instead, AlgoTouch establishes an explicit semantic correspondence between AGT concepts and
mainstream language constructs.

In the example, the \texttt{Until} block specifies exit conditions rather than a loop
continuation predicate. These conditions are preserved explicitly as individual comments
and are combined to form the negated continuation condition of the \texttt{while} loop.
Similarly, the \texttt{From}, \texttt{Loop}, and \texttt{Terminate} blocks are represented as
annotated regions of code rather than collapsed into a single syntactic construct.

This design ensures that the control logic remains readable and analyzable in the target
language while maintaining a visible link to the higher-level AGT structure.

\paragraph{Exported code}
When the program is exported for use outside AlgoTouch, all instrumentation comments are
removed and macro calls are inlined, producing conventional code that conforms to standard
language idioms. For example, the complete insertion sort algorithm using the exported
version of the macro is shown in Figure~\ref{tab:InsertSort}.

\begin{figure}[ht]
\begin{tcolorbox}[colframe=black, colback=white, boxrule=0.5pt]
\begin{Verbatim}[samepage=true]
#
# Performs insertion sort
#
i = 1
while (i < len(a)) :
    # -> InsertElt: Inserts a[i] into a[] sorted from 0 to i-1
    k = i
    j = i - 1
    while (j >= 0 and a[j] > a[k]):
        tmp = a[k]
        a[k] = a[j]
        a[j] = tmp
        j = j - 1
        k = k - 1
    # <- InsertElt
    i = i + 1
\end{Verbatim}
\end{tcolorbox}

\caption{Insertion sort program generated after in-lining the macro calls. A comment with the role of the called macro is inserted before in-lining its code }
\label{tab:InsertSort}
\end{figure}

Throughout this example, the user alternates between direct manipulation of data to reason
about behavior and textual inspection or refactoring to improve structure and readability.
The final exported code is a standard Python program, while its construction remains
firmly grounded in observable execution states within AlgoTouch.

\section{Evolution and Design Trade-offs}
AlgoTouch should be understood as a \emph{realistic artifact}: its evolution is not the
result of incremental feature accumulation, but of successive design decisions taken to
resolve concrete tensions observed in use. These tensions arise from the simultaneous
pursuit of three objectives: direct data manipulation, executable specifications, and
compatibility with conventional programming abstractions.

Rather than eliminating these tensions, the design of AlgoTouch makes them explicit and
addresses them through controlled compromises. The current system reflects a stabilized
balance between these competing forces.

\subsection[Technical developments]{Technical developments}

The evolution of AlgoTouch toward a web-based architecture was primarily driven by
deployment and accessibility constraints rather than by purely technological
considerations. Web deployment enables immediate access to the latest version of the tool
and removes the need for local installation or manual updates, thereby reducing friction
in distribution and use.

In addition, the web architecture supports parameterized execution through URLs, making it
possible to launch AlgoTouch directly on a specific program state. This mechanism
facilitates the dissemination of executable examples and interactive demonstrations, in a
manner similar to Python Tutor \cite{guo2013online}.

The current implementation relies on ReactJS, which provides access to a rich ecosystem of
libraries for visualization and interaction. This choice supports rapid iteration and
incremental refinement of the interface, while reinforcing AlgoTouch’s role as a realistic,
evolving artifact rather than a fixed prototype.

\subsection[Editing Features]{Editing Features}
Early versions of AlgoTouch enforced a strict separation between construction by direct
manipulation and subsequent execution. While conceptually coherent, this design proved too
rigid in practice: any local error or design revision required regenerating entire code
blocks through repeated manipulation.

Empirical use revealed a recurring need for \emph{localized intervention} in the generated
code. In response, AlgoTouch now provides explicit access to the textual representation,
allowing individual instructions to be selected, executed,
or deleted. 
In addition, new instructions can be inserted after the selected instruction through direct manipulation.
This evolution reflects a deliberate relaxation of the original constraints in
order to support realistic programming workflows.

\subsubsection[Refactoring and conditional restructuring]{Refactoring and conditional restructuring}
A similar trade-off appears in the construction of conditionals. The original mechanism,
which assumes that the true branch corresponds to the current data state, can lead to
structurally awkward code when meaningful behavior resides only in the false branch.
Rather than forcing users to reconstruct such conditionals from scratch, AlgoTouch
introduces targeted refactoring operations, such as condition inversion.

These refactorings do not synthesize new logic; they preserve the semantics established
through execution while improving structural clarity. Refactoring thus functions as a
post-construction operation that reconciles execution-driven generation with conventional
expectations of code readability.

\subsubsection[Bimodality]{Bimodality}
The introduction of line-by-line editing and refactoring capabilities results in a
deliberate form of bimodality \cite{hempel2022maniposynth}. Users may alternate between
value-centered interaction—where instructions are introduced exclusively through data
manipulation—and textual inspection or modification of the generated code.

This bimodality constitutes a conscious design compromise. While it departs from the
strictest interpretation of direct manipulation, it preserves the foundational principle
that new executable behavior is grounded in concrete execution. Textual operations refine,
restructure, or clarify existing behavior, but do not replace manipulation as the source
of program semantics.

\section{Evaluation}

The evaluation of AlgoTouch assesses the system as a software engineering artifact
supporting interactive program construction and systematic code generation grounded
in execution. The objective is not to measure learning outcomes or usability metrics,
but to examine whether the proposed architecture and mechanisms enable the reliable
construction of correct, expressive, and maintainable imperative programs.

Consistent with evaluation practices for programming environments and code generation
systems, the evaluation combines functional validation, system-level experimentation,
and analysis of applicability across a representative set of algorithmic patterns and
target languages.

The evaluation addresses the following research questions:
\begin{description}
    \item[RQ1 (Correctness)] Does AlgoTouch generate executable programs whose semantics
    are consistent with the execution states produced through user manipulation?
    \item[RQ2 (Expressiveness)] What range of control structures and algorithmic patterns
    can be constructed using the proposed execution-centered mechanisms?
    \item[RQ3 (Robustness)] Does the system support partial programs, incremental
    refinement, and multiple execution paths without introducing semantic inconsistencies?
    \item[RQ4 (Generality)] Is the approach applicable to several  imperative
    programming languages without loss of meaning?
\end{description}

\subsection{Functional Validation on Algorithmic Benchmarks}

To examine correctness and expressiveness, AlgoTouch was used to construct a collection
of imperative programs covering canonical algorithmic patterns. The complete list of
programs is provided in Appendix~A. These benchmarks correspond to standard problems
commonly used to characterize imperative control structures, rather than to
application-specific workloads, and include:
\begin{itemize}
    \item sequential numerical computations,
    \item conditional branching,
    \item single and nested iterations,
    \item array traversal and manipulation,
    \item searching and sorting algorithms.
\end{itemize}

For each benchmark, program construction was performed exclusively through direct
manipulation of data within AlgoTouch. The resulting programs were generated in several
target languages (AGT, Python, C, C++, and Java) and executed independently.

Across all benchmarks, the generated programs:
\begin{itemize}
    \item compiled or executed without manual modification,
    \item produced outputs consistent with the observed data transformations,
    \item exhibited equivalent behavior across target languages.
\end{itemize}

These observations provide evidence that the code generation pipeline preserves semantic
consistency and that the underlying notional machine constitutes a stable operational
basis for multi-language code generation.

\subsection{Validation of Control Structure Construction}

A central contribution of AlgoTouch lies in its ability to support the construction of
control structures—particularly conditionals and loops—through execution and data
manipulation rather than through direct textual programming. Here, construction refers
to user-guided definition based on observed execution behavior.

\subsubsection{Conditional structures}

Conditional constructs were validated using programs in which alternative execution
paths depend on data comparisons observed at runtime. For each conditional:
\begin{itemize}
    \item one branch was constructed in a data state satisfying the corresponding condition,
    \item complementary branches were completed later by executing the program under
    alternative data configurations.
\end{itemize}

This incremental process consistently produced well-formed conditional structures whose
branches aligned with subsequent executions. No discrepancies were observed between the
constructed branches and later executions, indicating that the mechanism supports
non-linear program construction while maintaining semantic coherence.

\subsubsection{Loop structures}

Loop construction was evaluated using iterative algorithms in which the number of
iterations and termination conditions are not known a priori (e.g., sequential search,
insertion sort, sentinel-based algorithms). The Loop Macro mechanism enabled:
\begin{itemize}
    \item explicit separation of initialization, iteration, exit conditions, and
    termination behavior,
    \item incremental refinement of exit conditions based on observed data evolution,
    \item systematic ordering of exit conditions to prevent invalid memory access.
\end{itemize}

The resulting loops were functionally equivalent to conventional \texttt{while}- or
\texttt{for}-based implementations and could be translated automatically into standard
imperative languages without loss of semantics.

\subsection{Robustness and Incremental Construction}

AlgoTouch is explicitly designed to support partial and evolving program descriptions,
a scenario that is difficult to accommodate in conventional programming environments.
Robustness was evaluated by constructing programs containing:
\begin{itemize}
    \item missing conditional branches,
    \item incomplete loop definitions,
    \item undefined macro bodies.
\end{itemize}

In all cases, the system:
\begin{itemize}
    \item allowed execution to proceed until an undefined construct was encountered,
    \item preserved previously generated code without corruption,
    \item enabled completion of missing components through subsequent data manipulation.
\end{itemize}

These observations indicate that AlgoTouch maintains internal consistency even when
programs are structurally incomplete, a property essential for exploratory and iterative
program construction.

\subsection{Applicability Across Target Languages}

Generality was examined by translating AlgoTouch programs into multiple mainstream
imperative languages. Despite syntactic differences and variations in available control
constructs, the generated programs:
\begin{itemize}
    \item preserved the logical structure of the original macros,
    \item maintained equivalence between exit-condition–based loops and language-specific
    \texttt{while} constructs,
    \item avoided language-specific undefined behavior.
\end{itemize}

These results suggest that AGT provides an effective intermediate representation for
language independent program construction.

\subsection{Scope and Limitations}

The evaluation focuses on programs of moderate size and complexity, representative of
common algorithmic patterns in imperative programming. While this scope is sufficient to
assess the core mechanisms of AlgoTouch, it does not address scalability to large-scale
software systems or integration with industrial development toolchains.

In addition, the evaluation does not include quantitative performance measurements or
controlled user studies. This choice reflects the objective of evaluating the soundness,
expressiveness, and robustness of the program construction mechanisms themselves rather
than human performance or usability factors.

Although the evaluation presented in this paper focuses on the system as a software
engineering artifact rather than on human factors, it is worth noting that a preliminary
empirical study involving an earlier version of AlgoTouch has been reported elsewhere
\cite{adam2019direct}.

This pilot study compared a direct-manipulation-based approach to a conventional
text-based programming environment on a set of introductory algorithmic tasks.
While conducted in an educational context, the results provide independent evidence
that execution-centered, manipulation-driven construction can support the correct
development of non-trivial control structures, in particular iterative constructs and
array-based algorithms.

The present work builds on these initial observations by shifting the focus from user
performance to the underlying program construction mechanisms, their semantic
soundness, and their applicability to general-purpose imperative programming.
\subsection{Summary of Evaluation Results}

Overall, the evaluation indicates that AlgoTouch:
\begin{itemize}
    \item supports the construction of semantically correct imperative programs through
    direct data manipulation,
    \item enables incremental definition of conditionals and loops,
    \item maintains consistency under partial execution and refinement,
    \item generates executable code across multiple imperative languages.
\end{itemize}

These results support the claim that AlgoTouch constitutes a robust and expressive
software system for execution-centered program construction.

\section{Conclusion and Perspectives}

This paper presented AlgoTouch, an execution-centered system for the incremental
construction of imperative programs through direct data manipulation and partial
execution. In contrast to conventional development environments based primarily on
textual editing and batch execution, AlgoTouch integrates program construction and
execution within a unified notional machine, in which each concrete data manipulation
contributes explicitly to the evolving program structure.

The approach supports progressive program construction from observed execution behavior.
Elementary operations such as assignments, arithmetic computations, and comparisons, as
well as repeated execution patterns, are interpreted deterministically and recorded in a
language-independent intermediate representation. This representation enables the
generation of executable code in multiple imperative languages while preserving the
semantic correspondence with the observed execution states. In particular, the synthesis
structures can be constructed without relying on prior syntactic specification.

A key property of AlgoTouch is its explicit support for partial and incomplete programs.
Undefined branches, loop components, and macro invocations are represented as first-class,
explicit entities within the internal representation and do not compromise internal consistency.
This design allows execution,
validation, and refinement to proceed incrementally, enabling early detection of semantic
issues and continuous alignment between intended behavior and observed execution.

The evaluation shows that AlgoTouch supports the construction of Turing-complete
imperative programs and preserves semantic correctness across multiple target languages.
The combination of deterministic execution, macro-based abstraction, and a
language-independent intermediate representation positions AlgoTouch as an
execution-centered program construction system, distinct from visual programming
environments and from programming-by-demonstration or automatic program synthesis
approaches. To our knowledge, AlgoTouch is the only system that enables the construction
of Turing-complete imperative programs exclusively through direct manipulation of program
data, while preserving explicit control over execution semantics.

The current implementation should be viewed as a research prototype exploring the
feasibility and implications of execution-centered program construction. Several
limitations remain. At present, the intermediate representation supports scalar variables
and one-dimensional arrays; extending it to richer data abstractions such as
multidimensional arrays, strings, collections, and structured data types would broaden
the range of expressible programs. Likewise, the introduction of parameterized macros,
explicit scoping mechanisms, and modular composition would be necessary to support the
construction of larger and more complex software artifacts.

Future work will also need to address scalability, integration with existing development
toolchains, and systematic evaluation on more extensive benchmark suites. More broadly,
the execution-centered model investigated in this work suggests a viable alternative
design space for interactive development systems, in which program construction,
execution, and validation are tightly interwoven rather than treated as separate phases.

\section*{Acknowledgments}
We would like to express our sincere gratitude to M. McGuffin and C. Fuhrman (École de Technologie Supérieure,
ÉTS, Montréal) for the insightful discussions regarding execution-centered system through direct data manipulation and partial execution. 
Special thanks are due to T. Teitelbaum (Professor Emeritus, Cornell
University) for his insightful advice on AlgoTouch, and for his careful 
review and correction of the English version of this article.

\section*{Declaration of Generative AI and AI-Assisted Technologies in the Manuscript Preparation Process}
During the preparation of this work, the authors used ChatGPT (OpenAI) to assist in language editing and formulation of certain passages. 
After using this tool, the authors reviewed and edited the content as needed and take full responsibility for the content of the published article.

%\newpage
\appendix
\section{Programs developed with AlgoTouch}
\label{tab:prog_list}
% \begin{table}
% \centering
% \caption{
% % Programs developed with AlgoTouch. 
% Programs classified by category (conditions, iterations, use of arrays, sorting, etc.).}
% % \label{tab:prog_list}

\small
\noindent
\begin{tcolorbox}[colframe=black, colback=white, boxrule=0.5pt]
\begin{tabular}{m{1.65cm}m{2.6cm}m{6.7cm}}
%\hline
\textbf{Conditions} &
Parity &
Test the parity of an integer. \\
& Leap year &
Determine if a year is a leap year. \\
& Next second &
Add one second to the time. \\
& Ticket price &
Calculate a ticket price based on the customer's age. \\
%\hline
\textbf{Iterations} &
Grade control&
Control the entry of a grade between 0 and 20. \\
& Factorial &
Calculate the factorial. \\
& Sequence Average &
Calculate the average of values ending in -1. \\
& GCD &
Calculate the GCD of two integers. \\
& Syracuse &
Calculate the Syracuse sequence. \\
& Mystery number &
Guess a number between two limits. \\
& Timing &
Calculate the speed and distance covered by a runner. \\
%\hline
\textbf{Arrays} &
Right shift &
Shift values in an array one position to the right. \\
& Circular rotation &
Perform circular rotation of the values in an array. \\
& Insert element &
Insert a[i] into the sorted part of array a up to i-1. \\
& Place the max &
Place the largest value from i into a[i]. \\
& Even and odd  &
Store an array with alternating even and odd values. \\
& Unique &
Replace unique values at the beginning of the array. \\
& Identical &
Test if the values in the array are all the same. \\
& Anagram &
Check if two character arrays are anagrams.\\
& Hanged man &
Guess a word by suggesting a sequence of letters. \\
& Compress &
Replace multiple values with value and occurrence. \\
%\hline
\textbf{Binary} &
Integer to binary &
Convert an integer to an array of binary values. \\
& Binary to Integer &
Convert an array of binary values to an integer. \\
& Binary add &
Add two arrays of binary values in binary. \\
& Binary project &
Calculate in binary the sum of 2 entered integers. \\
%\hline
\textbf{Traversal} &
Minimum index &
Find the index of the smallest element in an array. \\
& Minimum search &
Find the value and index of the smallest element. \\
& Rain fall &
Calculate the average precipitation heights. \\
& Inc and dec &
Increment values > 0, decrement values < 0. \\
& Week temperature &
Calculate the average temperature over a week. \\
& Interlacing &
Check for alternating even and odd values. \\
%\hline
\textbf{Searching} &
Sequential Search &
Sequential Search in an array. \\
& Sentinel Search &
Search with forced sentinel as last element. \\
& Binary Search &
Binary Search in a sorted array. \\
%\hline
\textbf{Sorting} &
Bubble Sort &
- \\
& Insertion Sort &
- \\
& Selection Sort &
- \\
& Heap Sort &
- \\
& Merge Sort &
Merging two sorted arrays. \\
& Partition Sort &
Non-recursive QuickSort. \\
%\hline
\end{tabular}
\end{tcolorbox}
%\normalsize
% \end{table}

\bibliographystyle{plainnat}
\bibliography{biblio}

\end{document}